\documentclass[prl,twocolumn,superscriptaddress,appendix]{revtex4}
\usepackage{graphicx,amsmath}

\def\bra#1{\mathinner{\langle{#1}|}}
\def\ket#1{\mathinner{|{#1}\rangle}}

\def\prjct#1{\mathinner{|{#1}\rangle}\!\!\mathinner{\langle{#1}|}}
\newcommand{\coh}[2]{\mathinner{|{#1}\rangle}\!\!\mathinner{\langle{#2}|}}

\def\text#1{\textrm{#1}}

\def\Ep{\mathcal{E}}
\def\be{\begin{equation}}
\def\ee{\end{equation}}

\begin{document}

\title{
A photon-pair source with controllable delay \\ based on shaped inhomogeneous broadening
of rare-earth doped solids}

\author{Pavel Sekatski}
\affiliation{Group of Applied Physics, University of Geneva, 1211 Geneva 4, Switzerland}

\author{Nicolas Sangouard}
\affiliation{Group of Applied Physics, University of Geneva, 1211 Geneva 4, Switzerland}

\author{Nicolas Gisin}
\affiliation{Group of Applied Physics, University of Geneva, 1211 Geneva 4, Switzerland}

\author{Hugues de Riedmatten}
\affiliation{Group of Applied Physics, University of Geneva, 1211 Geneva 4, Switzerland}
\affiliation{ICFO-Institute of Photonic Sciences, Mediterranean Technology Park, 08860 Castelldefels (Barcelona), Spain}
\affiliation{ICREA-Instituci\'o Catalana de Recerca i Estudis Avan\c{c}ats, 08015 Barcelona, Spain}

\author{Mikael Afzelius}
\affiliation{Group of Applied Physics, University of Geneva, 1211 Geneva 4, Switzerland}

\date{\today}

\begin{abstract}
Spontaneous Raman emission in atomic gases provides an attractive source of photon pairs with a controllable delay. We show how this technique can be implemented in solid state systems by appropriately shaping the inhomogeneous broadening. Our proposal is eminently feasible with current technology and provides a realistic solution to entangle remote rare-earth doped solids in a heralded way.
\end{abstract}
\maketitle

\begin{figure}
\includegraphics[width=5 cm]{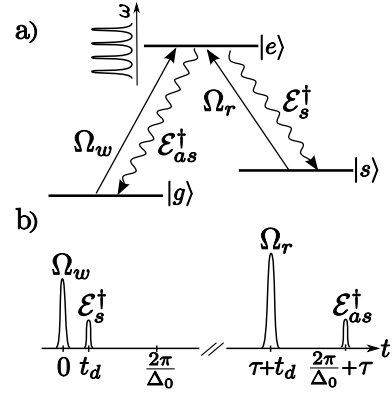}
\caption{a) Basic level scheme and b) pulse sequence for the creation of correlated Stokes -- anti-Stokes pairs via spontaneous Raman emission. All atoms starts in $g.$ The inhomogeneously broadened optical transition is shaped into an atomic frequency comb. A resonant write pulse $\Omega_w$ excites the $g$-$e$ transition, making possible the spontaneous emission of a Stokes photon. The detection of a Stokes photon (at time $t_d$) is uniquely correlated with the creation of a single collective atomic excitation corresponding to one excitation in $s$ delocalized over all the atoms. This collective excitation can be mapped very efficiently into an anti-Stokes photon using a resonant read pulse $\Omega_r$ (at time $\tau+t_d$) and beneficing from the echo-photon-type re-emission (at time $2\pi/\Delta+\tau$) inherent to inhomogeneous transition shaped into a frequency comb.}
\label{fig1}
\end{figure}

\paragraph{Introduction}
The creation of correlated Stokes -- anti-Stokes photon pairs in atomic ensembles via spontaneous Raman emission \cite{Duan01} plays a central role in quantum communication, starting with the implementation of quantum repeaters \cite{Sangouard09}. The basic principle requires an ensemble of lambda systems coupled to a pair of optical laser fields, see Fig. \ref{fig1}. An off-resonant laser pulse, the write pulse, produces a frequency shifted Stokes photon via spontaneous Raman emission. Detection of this Stokes photon in the far field, such that no information is revealed about which atom it came from, heralds the creation of a single collective atomic spin excitation. A remarkable feature of such a collective atomic state is that it can be read out very efficiently. A resonant laser pulse, the read pulse, allows one, through a collective spontaneous Raman emission, to ideally map the collective spin excitation into an anti-Stokes photon propagating in a well defined spatio-temporal mode.  This provides a photon-pair source with a very special property :  the delay between the Stokes and anti-Stokes photons can be controlled by choosing the timing between the write and read pulses. Such a source has inspired many experiments in atomic gases, including the first single-photon storage in an atomic ensemble \cite{Chaneliere05, Eisaman05}, the first heralded creation of entanglement between atomic ensembles \cite{Chou05} and even the implementation of the first elementary blocks of quantum repeaters \cite{Chou07}. Despite this impressive body of work, there are strong motivations to use more practical systems, e.g. in the solid state. Rare-earth-doped solids seem naturally well suited, at least at first sight. They are widely available thanks to their use for solid-state lasers. Thanks to their particular electronic structure, they can be seen as a frozen gas of atoms, with optical and spin transitions featuring excellent coherence properties. Moreover, they have already shown excellent capability to store light for long times \cite{Longdell05} with high efficiency \cite{Hedges10} and negligible noise \cite{deRiedmatten08, Hedges10}. Last but not least, they have large inhomogeneous absorption spectrum and narrow homogeneous lines leading to a high temporal multimode capacity \cite{Usmani10}.\\

All rare-earth elements have in common weak dipole moments on the relevant 4f - 4f transitions and large inhomogeneously broadened spectra due to the interaction with the host crystal. These two properties make the creation of Stokes -- anti-Stokes pairs in rare-earth doped solids challenging. In hot alkali gases, where there is also an inhomogeneous broadening due to the Doppler effect, spontaneous Raman processes have been successfully performed \cite{Eisaman05} with a write pulse far detuned from the resonance -- the probability $p$ for the emission of a Stokes photon in a given mode being enhanced by the use of large write intensities. However, in rare-earth doped solids, where the dipole moments are typically two to three orders of magnitude weaker, the required intensities would be, at best, difficult to achieve. For resonant write pulses, atoms are transfered into the excited state, leading to a higher $p$. But their energy difference, due to the inhomogeneous broadening, makes them distinguishable. They can no more interfere, which makes the readout of the collective excitation inefficient \cite{Ottaviani09}.\\
We propose a simple solution to this problem. It consists in shaping the spectral inhomogeneous broadening of the optical transition so that the atomic dipoles rephase at the readout step, leading to an efficient emission of the anti-Stokes photon even when resonant write pulses are used. Several shaping methods are available to force the atomic dipoles to rephase. For example, the inhomogeneous broadening could be shaped into a narrow absorption line with a reversible and controllable broadening \cite{crib}. In what follows, we focus on a shaping based on a comb-like structure composed with periodic narrow peaks  \cite{Afzelius09}. This provides a temporally multiplexed version of a spontaneous Raman source, similarly to the proposal of Ref. \cite{Simon10} in atomic gases but without the need for a cavity. Since spontaneous Raman based protocols are well suited for entangling remote atomic ensembles in a heralded way \cite{Duan01} without the need for ultra-narrowband pair sources, our proposal paves the way to the implementation of the first elementary link of quantum repeaters with solid-state devices.\\

\paragraph{Write step} Let us start by a description of the write step. We consider a medium consisting of lambda atoms (as depicted in Fig. \ref{fig1}) initially prepared in the state $g.$ The optical transition $g$-$e$ is shaped into a frequency comb made of narrow peaks with a characteristic width $\gamma,$ separated by $\Delta_0$ and spanning a large atomic frequency range $\Gamma.$ The $g$-$s$ transition is considered to be homogeneous. A weak write laser pulse with the Rabi frequency $\Omega_w(t),$ is sent through the medium to transfer a small part of atoms in the excited state $e$ in such a way that it can spontaneously decay into the level $s$ by emitting Stokes photons. Taking the absorption into account, the write pulse drives the atoms into
$$
\ket{\psi_w}=\prod_{j\geq 1}^N ( \underbrace{\frac{1}{n_j}(1-\frac{1}{2}\theta_0^2 e^{-\bar \alpha z_j})}_{:=G_j}\ket{g_j}+\underbrace{e^{i k_w z_j}\frac{\theta_0 e^{-\bar \alpha z_j/2}}{n_j}}_{:=E_j}\ket{e_j}).
$$
$n_j$ ensures the normalization, $\!\theta_0= \frac{1}{2}\int ds\times$  $\Omega_w(s)~\ll~1$ refers to the area of the write pulse at the entrance of the crystal and $k_w$ corresponds to its wave number. $z_j$ is the position of the $j$-th atom within a medium of length $L$ and $N$ is the total number of atoms. $\bar \alpha$ is the absorption per unit length of the atomic medium.  It is proportional to the ratio between the absorption per peak $\alpha$ and the finesse $F$ of the comb, c.f. below. For concreteness, we consider an atomic distribution made of gaussian peaks with full width half maximum $\gamma$ so that  $\bar \alpha = \frac{\alpha}{F}\sqrt{\frac{\pi}{4 \ln 2}}$ where $F=\Delta_0/\gamma$ (see appendix). Note that this shaping, which is at the heart of memories based on atomic frequency comb \cite{Afzelius09}, is known to produce a photon echo-type of reemission in a well defined spatial mode at time $2\pi/\Delta_0.$ Here, on the other hand, we look at the spontaneous emission of a Stokes photon at time $t_d < 2\pi/\Delta_0$ and we benefit from the efficiency of the photon-echo-like remission for the readout of the atomic spin wave.\\

\paragraph{Spontaneous emission of Stokes photons}
Let us consider the Stokes field which propagates in the forward direction with the carrier frequency $\omega_s.$ Its envelop is described by the slowly varying quantum operator \cite{note_freefield}
\be\label{Def:Envelop}
\hat\Ep_s(z,t) = \sqrt{\frac{L}{2\pi c}}e^{i\omega_s(t-z/c)}\!\int d\omega
\hat a_\omega(t) e^{i \omega z/c}.
\ee
Under the dipole and rotating wave approximation, the interaction between the Stokes field and the medium is governed by the Hamiltonian
\be\label{Def:Coupling}
H_\text{int} = -\hbar g\sqrt{\frac{L}{2\pi c}}\sum_j \int d\omega \hat a_\omega e^{i \omega z_j/c} \coh{e}{s}^j +\text{h.c.}
\ee
$g = \wp \sqrt{\frac{ \omega_s}{2 \hbar \varepsilon_0 A L}}$ is the atom-field coupling constant with $\wp$ the dipole moment of the $e - s$ transition and $A$ the interaction section. The equations of motion for the Stokes field and for the atomic coherence $\sigma_{se}^j(t) = \coh{s}{e}^j e^{i \omega_s (t-z_j/c)}$ are given by
\begin{align}\label{motionF}
(\frac{d}{dt}+c\frac{d}{dz})\hat\Ep_s(z,t) = i g  L \sum_j \delta(z-z_j)\sigma_{se}^j
\\\label{motionA}
\frac{d}{dt}\sigma_{se}^j = - i \Delta^j\sigma_{se}^j - i g \hat \Ep_s(z_j,t)
\sigma_{z, s}^j.
\end{align}
$\Delta^j = \omega_{es}^j-\omega_s$ is the frequency detuning and $\sigma_{z, s}^j =\prjct{e}^j- \prjct{s}^j$ is approximated by its mean value $\theta_0^2 e^{-\bar \alpha z_j}$ in what follows. Plugging the formal solution of equation (\ref{motionA}) into equation (\ref{motionF}), we obtain the expression of the Stokes field at $z$ (see appendix)
\begin{eqnarray}\label{StokesEmission}
& \hat\Ep_s(z,t)& = e^{ \frac{\bar \alpha}{2}\int_0^z\theta_0^2e^{-\bar \alpha z'}dz'} \hat\Ep_s(0,t)\\
 \nonumber
 &&+i \frac{g  L}{c} \sum_{j | z_j < z}e^{ \frac{\bar \alpha}{2}\int_{z_j}^z\theta_0^2 e^{-\bar \alpha z'}dz'}e^{-i\Delta^jt}\sigma_{se}^j(0).
\end{eqnarray}
We considered for simplicity, that the transitions $g$-$e$ and $s$-$e$ have the same dipole moments. Since the state of the complete system after the write pulse is given by $|\Psi_w\rangle=|\psi_w,0\rangle$ where $|0\rangle$ is the vaccum for the electromagnetic field, the average number of Stokes photons emitted at time $t_d$ in a mode of temporal duration $\sqrt{2\pi}/\Gamma$ is given by
\begin{equation}
\label{StockesProb}
\frac{\sqrt{2\pi}c}{\Gamma L}\bra{\Psi_w}\hat \Ep_s^\dag (L,t_d) \hat \Ep_s(L,t_d)\ket{\Psi_w} \approx \theta_0^2(1-e^{-\bar \alpha L})
\end{equation}
for $\theta_0 \ll 1$ (see appendix). This formula is a very useful and can easily be used in practice. Note first that the relative number of atoms transferred into the excited state by the write pulse is given by $\frac{1}{N}\int_0^{L} \theta_0^2 e^{-\bar \alpha z}\frac{Ndz}{L}=\frac{\theta_0^2}{\bar \alpha L}(1-e^{-\bar \alpha L}).$ Therefore, the formula (\ref{StockesProb}) tells us that the average number of photons in a mode with a duration corresponding to the inverse of the overall spectrum, is merely the optical depth $\bar \alpha L$ times the relative number of atoms in the state $e.$ In what follows, we are interested in the regime where the optical depth $\bar\alpha L$ is large to get high readout efficiencies but the write pulse is weak to get a high signal-to-noise ratio for the readout, c.f. below. In this case, the success probability for the emission of a Stokes photon (\ref{StockesProb}) reduces to $p\approx \theta_0^2$. \\

\paragraph{Readout efficiency} We now calculate the efficiency of the readout process. At time $\tau$ after the detection of the Stokes photon, a read pulse resonant with the transition $s$-$e$ and associated with the Rabi frequency $\int ds \Omega_r(s)=\pi$ goes through the atomic ensemble and exchanges the population of states $s$ and $e.$ For $\theta_0 \ll 1,$ the resulting atomic state is given by (see appendix)
\begin{align}
&|\psi_{r}\rangle=\zeta \sum_{j\ge 1}^N  e^{i z_j (k_w-k_r-\omega_s/c)}e^{-\bar \alpha z_j/2} e^{-i\Delta_j t_d} \ket{e_j} \times
\label{cond_state}  \\
\nonumber
&\prod_{\ell \ne j} \left(G_\ell  e^{-i \omega_{gs} (t_d+\tau)}\ket{g_\ell}+E_\ell e^{i z_\ell k_r}e^{-i \omega_{es}^{\ell} (t_d+\tau)}\ket{s_\ell}\right)
\end{align}
with $\zeta=(\bar \alpha L)^{\frac{1}{2}}/(N(1-e^{-\bar \alpha L}))^{\frac{1}{2}}.$ Consider a re-emission associated with an anti-Stokes mode $\hat\Ep_{as}(z,t) $ propagating in the backward direction. Following the method presented before (under the assumption that $\prjct{e}^j- \prjct{g}^j \approx -1$) we find at $z=0$
\begin{align}
\label{antiStokesEmission}
&\hat\Ep_{as}(0,t) = e^{- \frac{\bar \alpha}{2}L} \hat\Ep_{as}(L,t) \nonumber\\
&+i \frac{g  L}{c} \sum_{z_j }e^{- \frac{\bar \alpha}{2}z_j}e^{-i\Delta^j(t-(\tau+t_d))}\sigma_{ge}^j(\tau+t_d)
\end{align}
so that, at time $2\pi/\Delta+\tau$ where all the atoms are in phase, the efficiency of the readout process is
\begin{eqnarray}
\label{AntiStokesProbability}
\nonumber
\frac{\sqrt{2\pi}c}{\Gamma L} &  \bra{\Psi_r} \hat \Ep_{as}^\dag(0,2\pi/\Delta+\tau) \hat \Ep_{as}(0,2\pi/\Delta+\tau)\ket{\Psi_r} & \\
& = (1-e^{-\frac{\alpha L}{F}\sqrt{\frac{\pi}{4 \ln 2}}}) \-\ e^{-\frac{\pi^2}{2\ln 2 \-\ F^2}}.&
\end{eqnarray}
One sees that there is a tradeoff between absorption and dephasing. However, for large enough $\alpha L$ and optimized $F,$ the readout efficiency can be arbitrary close to 100\%. Let us directly note some of the advantages of spontaneous Raman protocol over other schemes. First, the proposal based on spontaneous Raman is significantly more efficient than a memory where the photon has first to be absorbed before being reemitted. It leads to a much higher efficiency for small optical depths (see appendix). Furthermore, since the readout of spontaneous Raman protocols is conditioned on the detection of a Stokes photon, the retrieved signal is not affected by the non-unit coupling efficiency of an input photon into the memory (e.g. due to imperfect spectral filters, non-unit coupling into monomode fibers $\hdots)$ (see appendix). Contrary to the photon-pair source based on rephased amplified spontaneous emission \cite{Ledingham10}, our proposal provides highly correlated pairs even for large optical depths where the retrieval efficiency is high. \\

\paragraph{Noise rate} We now account for intrinsic noise. The conditional state (\ref{cond_state}) that we used to calculate the efficiency of the readout process corresponds to the ideal case where a single Stokes photon has been emitted and detected. However, many atoms prepared in the excited state by the write pulse, can emit Stokes photons in all the spatio-temporal modes. This unwanted emission populates the state $s$ so that after the interaction with the read pulse, atoms occupy the excited state $e$ and can produce spontaneous noise in the anti-Stokes mode. To take this noise into account, we consider the worst case where all the atoms prepared in the excited state by the write pulse decay on the $e$-$s$ transition. Tracing over the Stokes photons, the atomic state after the interaction with the read pulse is well approximated by
\begin{equation}
\varrho_n= \bigotimes_{j\geq 1}^N ( (1-\theta_0^2 e^{-\bar \alpha z_j})\coh{g_j}{g_j}+ \theta_0^2 e^{-\bar \alpha z_j}\coh{e_j}{e_j} ).
\end{equation}
The noise is then deduced from
\begin{align}
\nonumber
\frac{\sqrt{2\pi}c}{\Gamma L}\text{tr} \big(\hat \Ep_{as}^{\dagger}(0,2\pi/\Delta+\tau)  \hat \Ep_{as}(0, & 2\pi/\Delta+\tau) \rho_n \big) \\
\nonumber
&=  \frac{\theta_0^2}{2} (1- e^{-2\bar\alpha L })
\end{align}
where $\rho_n=\varrho_n \otimes |0\rangle\langle 0|.$ For large enough $\alpha L $ and optimized $F$, the signal-to-noise ratio is lower bounded by
\begin{equation}
\text{signal-to-noise} \geq \frac{2}{\theta_0^2} = \frac{2}{p}.
\end{equation}
One can thus conclude that despite inhomogeneous broadening inherent to solid-state systems, our protocol achieves similar characteristics to the ones obtained in cold atomic gases - the signal-to-noise can be very high provided that the number of Stokes photons per mode is low. \\

\paragraph{Feasibility} For concretness, we now discuss the experimental feasibility of spontaneous Raman processes in rare-earth-doped materials. Pr:Y$_2$SiO$_5$ is a very promising material for initial experiments, since excellent hyperfine coherence \cite{Fraval05} and quantum memory efficiencies of order 70\% \cite{Hedges10} have already been demonstrated. The main drawback of Praseodymium is the small hyperfine separation (a few MHz) which limits the number of peaks within the atomic comb and thus the multimode capacity, c.f. below. If we consider the $^3$H$_4$ - $^1$D$_2$ transition at 606 nm, one can realistically shape combs with individual peaks of FWHM $\gamma \approx 30$kHz \cite{Hetet08} over a spectral range of $\Gamma \approx 2$MHz and an absorption per peak of order $\alpha L \approx 10$  \cite{Amari10} (the branching ratio is approximated by 1). For F=5, $p\approx 0.1$ can be achieved provided that the write pulse satisfy $\theta_0^2=0.1$ at the entrance of the crystal. This would lead to a readout efficiency of $65\%$ and a signal-to-noise ratio larger than 10 under the assumption that the noise is dominated by the spontaneous emission. Now consider a setup involving two Pr-doped solids located 1 km apart so that the corresponding Stokes modes are combined on a beamsplitter at a central station. Further consider a fiber attenuation of 9 dB/km, corresponding to 606 nm photons \cite{Takesue10} and assume a coupling efficiency into optical fibers of $\eta_c=50\%.$ The average time to detect a Stokes photon after the beamsplitter and thus to entangle the two crystals is $T=\frac{1}{2rp\eta_c\eta_t\eta_d}$ where $\eta_t,$ $\eta_d$ are the transmission and detection efficiencies and $r$ is the repetition rate. Assume $\eta_d=70\%.$ For $p=0.05$ where the fidelity of the entanglement is $\mathcal{F}=1-3p(1-\eta_c\eta_t\eta_d) \approx 0.85$ \cite{Sangouard09}, one finds $T \approx 0.1$s for a repetition rate of $r=1$KHz so that a few hours would be sufficient for performing a full tomography. Note that the fiber lengths have to be actively stabilized to guarantee an interferometric phase stability on this time scale \cite{Sangouard09}. \\

\paragraph{Conclusion} Our approach opens an avenue towards the heralded entanglement of remote solids. Beyond that, the present scheme might be useful for quantum communications since spontaneous Raman processes lead to the production of narrowband pairs well suited for storage in atomic ensembles. Let us finally emphasize that in our protocol, spin waves created at different times, say $t_{d1},t_{d2},t_{d3},\hdots$ are independent and if the read pulse is sent at time $\tau$ after the first detection, these spin waves rephase at times $\tau+2\pi/\Delta, \tau+2\pi/\Delta-(t_{d2}-t_{d1}), \tau+2\pi/\Delta-(t_{d3}-t_{d1}), \hdots.$ The number of spin waves that can be stored is roughly given by the number of peaks composing the comb (see \cite{Afzelius09} and appendix) and can be merely increased by making use of wider range of the inhomogeneous broadening. In the framework of quantum repeaters, this temporal multiplexing has been shown to greatly enhance the distribution rate of entanglement \cite{Simon07}.\\

We thank T. Chaneliere, J.-L. Le Gouet, J. Minar and C. Simon for helpful comments and interesting
discussions. We gratefully acknowledge support by the EU project {\it Qurep} and the Swiss NCCR {\it Quantum Photonics}.


\vspace{15 pt}
\appendix
\begin{center}\textbf{APPENDIX}
\end{center}
\vspace{5 pt}
\paragraph{Atomic distribution}
For concreteness, we consider an atomic spectral distribution shaped into a frequency comb made with gaussian peaks
\be\label{dist}
\Theta(\Delta) := \frac{\Delta_0}{2\pi \tilde \gamma \Gamma} e^{-\frac{\Delta^2}{2\Gamma^2}} \sum_{j=-\infty}^\infty e^{-\frac{(\Delta + j \Delta_0)^2}{2\tilde \gamma^2}}.
\ee
$\Gamma$ denotes the overall width, $\Delta_0$ is the peak separation and $\tilde\gamma$ is the width of an individual peak. \\
We are interested in the regime $\Gamma \gg \Delta_0 \gg \tilde\gamma $ where a large number of well separated peaks spans a large spectral bandwidth. In this regime, one can check that $\int d\Delta \Theta(\Delta) = 1.$ \\
We define the finesse of the comb as the ratio of the peak separation over the full width at half-maximum of a single peak $\gamma,$ i.e.
$$
F :=\frac{\Delta_0}{\gamma}=\frac{1}{\sqrt{8 \ln 2}}\frac{\Delta_0}{\tilde \gamma}.
$$
Note that the Fourier transform of the atomic spectral distribution
$$
\tilde\Theta(t):=\int d\Delta \Theta(\Delta) e^{-i\Delta t}
$$
is also a series of gaussian peaks with individual peak width $1/\Gamma,$ separated by $2\pi/\Delta_0$ and spanning the overall temporal width $1/\tilde\gamma.$ \\
Further note that we consider a perfect three-level system (without additional levels) and we assumed that the optical transitions $g-e$ and $e-s$ have the same dipole moments. Therefore, the absorption of $g-e$ with all the atoms in $g$ is the same as the one associated to $e-s$ if all the atoms are prepared in $s.$\\

\paragraph{Emission of Stokes photons}
We now detail the way to find the solution of equations associated to the dynamics of the Stokes field
\begin{align}\label{motionF}
(\frac{d}{dt}+c\frac{d}{dz})\hat\Ep_s(z,t) = i g  L \sum_j \delta(z-z_j)\sigma_{se}^j
\\\label{motionA}
\frac{d}{dt}\sigma_{se}^j = - i \Delta^j\sigma_{se}^j - i g \hat \Ep_s(z_j,t)
\sigma_{z, s}^j.
\end{align}
First, we plug the formal solution of the equation (\ref{motionA})
\be
\label{sigma_se}
\sigma_{se}^j(t) =  -i g \,\sigma_{z, s}^j \int_{0}^t dt'e^{-i\Delta^j(t-t')} \hat \Ep_s(z_j,t') + e^{-i\Delta^jt}\sigma_{se}^j(0).
\ee
into the equation (\ref{motionF}) and since we consider Stokes modes with a characteristic duration $\tau_s \gg L/c$, we can neglect the temporal retardation effects in the crystal, i.e. the time derivative in the equation (\ref{motionF}). This leads to
\begin{align}
\label{field}
\frac{d}{dz}\hat\Ep_s(z,t) = i \frac{g L}{c} \sum_j \delta(z- z_j) e^{- i \Delta^j t} \sigma_{se}^j (0)+
\nonumber\\
 \frac{g^2 L}{c} \sum_j \delta(z-z_j) \sigma_{z,s}^j \int_0^t dt' e^{-i \Delta^j(t-t')}\hat\Ep_s(z_j,t').
\end{align}
We then replace $\sigma_{z,s}^j$ by its mean value calculated on the state $\ket{\psi_w},$ i.e. $\sigma_{z,s}^j\approx \theta_0^2 e^{-\bar \alpha z_j}$ and we take the continuous limit $\sum_j \rightarrow \int_0^L \frac{N}{L}dz' \int_{-\infty}^\infty d\Delta \Theta(\Delta),$ $N$ being the total number of atoms and $L$ the length of the medium. The last term of the equation (\ref{field}) reduces to
$$
\frac{g^2 N}{c} \theta_0^2 e^{-\bar \alpha z} \int_0^t dt' \tilde\Theta(t-t')~\hat\Ep_s(z,t')
$$
Since we are interested in the emission process which has a typical duration $\tau_s,$ we only need to consider values of $t-t'$ of order $\tau_s.$ Considering the regime where $\tau_s \ll  2\pi/\Delta_0,$ only the central peak of $\tilde\Theta$ contributes and $\tilde\Theta(t-t')$ is well approximated by
$$
\tilde\Theta(t-t') \approx e^{-(t-t')^2 \Gamma^2/2}.
$$
For $\Gamma \tau_s > 1,$ $\tilde\Theta(t-t')$ acts as a delta function
$$
\tilde\Theta(t-t') \approx e^{-(t-t')^2 \Gamma^2/2} \approx \frac{\sqrt{2\pi}}{\Gamma}\delta(t-t')
$$
and the equation for the Stokes field reduces to
\begin{align}
\frac{d}{dz}\hat\Ep_s(z,t) = i \frac{g L}{c} \sum_j \delta(z- z_j) e^{- i \Delta^j t} \sigma_{se}^j (0)
\nonumber
\\
\label{eq6}
+\frac{\bar \alpha}{2} \,\theta^2_0 e^{-\bar \alpha z} \hat\Ep_s(z,t).
\end{align}
$\bar \alpha$ which corresponds to the optical depth per unit length, is defined by
\begin{equation}
\bar \alpha:=\frac{\sqrt{2\pi}}{c} \frac{g^2N}{\Gamma}.
\end{equation}
In the main text, we introduced the optical depth per unit length associated to the central peak $\alpha.$ It is given by $\alpha:=\frac{g^2N}{c}\frac{\Delta_0}{\tilde\gamma \Gamma}$ so that $\bar \alpha = \sqrt{\frac{\pi}{4\ln 2}}\frac{\alpha}{F}.$\\
It is then straightforward to check that the solution of the equation (\ref{eq6}) is given by
\begin{eqnarray}\label{StokesEmission}
& \hat\Ep_s(z,t)& = e^{ \frac{\bar \alpha}{2}\int_0^z\theta_0^2e^{-\bar \alpha z'}dz'} \hat\Ep_s(0,t)\\
 \nonumber
 &&+i \frac{g  L}{c} \sum_{j | z_j < z}e^{ \frac{\bar \alpha}{2}\int_{z_j}^z\theta_0^2 e^{-\bar \alpha z'}dz'}e^{-i\Delta^jt}\sigma_{se}^j(0).
\end{eqnarray}
To find the average number of Stokes photons per temporal mode, we first evaluate the ket $\hat\Ep_s(L,t_d)\ket{\Psi_w}.$ Since initially, there is no excitation in the Stokes mode $|\Psi_w\rangle=|\psi_w,0\rangle,$ we obtain
\begin{eqnarray}
\nonumber
&\hat\Ep_s(L,t_d)&\ket{\Psi_w}=i\frac{g L}{c} \\
\nonumber
&&  \times \sum_j \left(e^{\frac{\bar \alpha}{2}\int_{z_j}^L \theta_0^2 e^{-\bar \alpha z'}dz'}e^{-i\Delta^j t_d} e^{-i \omega_s z_j/c}E_j\ket{s_j}\right.
\nonumber \\
&&\left.\times \prod_{k\neq j} (G_k \ket{g_k}+ E_k \ket{e_k})\right)\ket{0}.
\end{eqnarray}
Note that each term of the sum has only one atom in the state $\ket{s_j}$, so when taking the scalar product with the corresponding bra, only the term with the same atom in $\bra{s_j}$ will give a non zero contribution. Under the approximation $|E_j|^2 \approx \theta_0^2 \,e^{-\bar \alpha z_j},$ one gets
\begin{align}
\bra{\Psi_w}\hat\Ep_s^\dag(L,t_d)\hat\Ep_s(L,t_d)&\ket{\Psi_w}= \nonumber
\\ \frac{g^2 L^2}{c^2}\sum_j & e^{\bar \alpha\int_{z_j}^L \theta_0^2 e^{-\bar \alpha z'}dz'} \theta_0^2 \,e^{-\bar \alpha z_j}.
\end{align}
Finally, we replace the sum over $j$ by the integral expression. Since there is no term with a spectral dependence, the integral over $\Delta$ is carried out directly and gives one. This yields to the spatial density of photons
\begin{eqnarray}
&& \frac{1}{L}\bra{\Psi_w}\hat\Ep_s^\dag(L,t_d)\hat\Ep_s(L,t_d)\ket{\Psi_w} \nonumber
\\
&& =\frac{g^2 N}{c^2}\int_0^L \!\!dz\,  e^{\bar \alpha\int_{z}^L \theta_0^2 e^{-\bar \alpha z'}dz'} \theta_0^2 \,e^{-\bar \alpha z}  \nonumber \\
\nonumber
&& = \frac{g^2 N}{\bar \alpha c^2} (e^{\bar \alpha \int_0^L \theta_0^2 e^{-\bar \alpha z'}dz'}-1) \\
\nonumber
&& = \frac{g^2 N}{\bar \alpha c^2} (e^{ \theta_0^2 (1-e^{-\bar \alpha L})}-1) \\
\nonumber
&& \approx \frac{\Gamma}{\sqrt{2\pi} c} \theta_0^2 (1-e^{-\bar\alpha L})
\end{eqnarray}
where, at the last line, we expanded the exponential to the first order. Consequently, the number of photons per temporal mode of duration $\sqrt{2\pi}/\Gamma$ is given by
\begin{equation}
\frac{\sqrt{2\pi}}{\Gamma}\frac{c}{L} \bra{\Psi_w}\hat\Ep_s^\dag(L,t_d)\hat\Ep_s(L,t_d)\ket{\Psi_w} \approx \theta_0^2 (1-e^{-\bar \alpha L}).
\end{equation}
Note that in the general case, the number of photons emitted in a temporal mode $\sqrt{2\pi}/\Gamma$ is given by $e^{\bar \alpha \int_0^L \langle \sigma_{z,s}  \rangle  dz}-1.$ In the particular case where all the atoms are prepared in $e,$ the average number of Stokes photons per mode is given by $e^{\bar \alpha L}-1.$ This agrees with the result presented in Ref. \cite{Ledingham10}.\\
Further note that the number of Stokes photons emitted per write attempt is given by
\begin{equation}
\frac{2\pi}{\Delta_0}\frac{c}{L} \bra{\Psi_w}\hat\Ep_s^\dag(L,t_d)\hat\Ep_s(L,t_d)\ket{\Psi_w} \approx \frac{\sqrt{2\pi}\Gamma}{\Delta_0}\theta_0^2 (1-e^{-\bar \alpha L})
\end{equation}
and is thus roughly the product of the number of peaks $\Gamma/\Delta_0$ composing the atomic frequency comb by the success probability for the emission of a Stokes photon per mode. It can thus be merely increased by making use of wider range of the inhomogeneous broadening.\\

\paragraph{Atomic state prepared by the Stokes photon detection}
Consider the successful event where a Stokes photon is detected at time $t_d.$ The conditional atomic state $\psi_{d}$ is obtained by calculating the evolution of the initial state $\Psi_w$ until the time $t_d$
and by projecting the resulting state into $\bra{0} \hat \Ep_s (L,t_d).$ First, let us directly note that the initial state can be written as
$
|\Psi_w\rangle=\prod_{j\geq 1}^N \left( G_j \sigma_{gs}^j(0)+E_je^{-i \omega_s z_j/c}\sigma_{es}^j(0)\right)\ket{s_j}|0\rangle
$
where  $\sigma_{gs}^j(0) = \coh{g_j}{s_j}.$ Since the state $|s_j\rangle |0\rangle$ is an eigenstate associated to the eigenvalue 0, it stays unchanged in time. Furthermore, the operator $\sigma_{gs}^j$ merely acquires the phase term $e^{-i \omega_{gs} t_d}.$ The evolution of $\sigma_{es}^j$ is obtained by plugging the field solution (\ref{StokesEmission}) back into the atomic solution (\ref{sigma_se}) and then taking the adjoint. For $\theta_0 \ll 1,$ we get
\begin{eqnarray}
\nonumber
|\psi_{d}\rangle&=&\zeta \sum_{j\ge 1}^N e^{i z_j (k_w-\omega_s/c)}e^{-\bar \alpha z_j/2} e^{-i\Delta_j t_d}\ket{s_j} \\
\nonumber
&& \times \prod_{\ell \ne j} \left( G_\ell e^{-i \omega_{gs}t_d}\ket{g_\ell}+E_\ell e^{-i \omega_{es}^{\ell} t_d} \ket{e_\ell}\right)
\end{eqnarray}
with $\zeta=\frac{(\bar \alpha L)^{\frac{1}{2}}}{(N(1-e^{-\bar \alpha L}))^{\frac{1}{2}}}.$ Furthermore, at time $\tau$ after the detection of the Stokes photon, a read pulse resonant with the transition $s$-$e$ and associated with the Rabi frequency $\int ds \Omega_r(s)=\pi$ goes through the atomic ensemble and exchanges the population of states $s$ and $e.$ Therefore, the atomic state becomes
\begin{align}
&|\psi_{r}\rangle=\zeta \sum_{j\ge 1}^N  e^{i z_j (k_w-k_r-\omega_s/c)}e^{-\bar \alpha z_j/2} e^{-i\Delta_j t_d} \ket{e_j} \times
\nonumber   \\
&\prod_{\ell \ne j} \left(G_\ell  e^{-i \omega_{gs} (t_d+\tau)}\ket{g_\ell}+E_\ell e^{i z_\ell k_r}e^{-i \omega_{es}^{\ell} (t_d+\tau)}\ket{s_\ell}\right).
\nonumber
\end{align}
We now have all the necessary ingredients to calculate the efficiency of the readout process.\\

\paragraph{Readout efficiency}
The operator corresponding to the envelop of the anti-Stokes field propagating in the backward direction
is given by
\be
\hat\Ep_{as}(z,t) = \sqrt{\frac{L}{2\pi c}}e^{i\omega_{as} (t+z/c)}\!\int d\omega
\hat a_\omega(t) e^{i \omega z/c}.
\ee
Its evolution is given the equations of motion
\begin{align}\label{motionFas}
(\frac{d}{dt}-c\frac{d}{dz})\hat\Ep_{as}(z,t) = - i g  L \sum_j \delta(z-z_j)\sigma_{ge}^j,
\\\label{motionAas}
\frac{d}{dt}\sigma_{ge}^j = - i \Delta^j\sigma_{ge}^j + i g \hat \Ep_{as}(z_j,t)
\sigma_{z, g}^j
\end{align}
where $\sigma_{z, g}^j=\prjct{e}^j- \prjct{g}^j $ is approximated by $\sigma_{z, g}^j\approx -1$ in what follows. Using the methods presented before, we find at $z=0$
\begin{align}
\label{antiStokesEmission}
\hat\Ep_{as}(0,t) = e^{- \frac{\bar \alpha}{2}L} \hat\Ep_{as}(L,t) \nonumber\\
+i \frac{g  L}{c} \sum_{z_j }e^{- \frac{\bar \alpha}{2}z_j}e^{-i\Delta^j(t-(\tau+t_d))}\sigma_{ge}^j(\tau+t_d).
\end{align}
This solution allows one to evaluate the readout efficiency. For a perfect spatial phase matching (all the wave vectors sum to zero) the ket $\hat \Ep _{as}(0,t)\ket{\Psi_r}$ is given by (up to a global phase factor)
\begin{align}
i \frac{ g L}{c}  \zeta \sum_j e^{-\bar \alpha z_j} e^{-i\Delta^j(t-\tau)}\ket{g_j}\times \nonumber\\
\prod_{\ell \ne j} \left( G_\ell e^{-i \omega_{gs}(t_d+\tau)}\ket{g_\ell}+E_\ell e^{i z_\ell k_r}e^{-i \omega_{es}^{\ell} (t_d+\tau)}\ket{s_\ell}\right)\ket{0}.
\end{align}
Projecting on the corresponding bra, one can show that the spatial density of photons is well approximated by
\begin{align}
\frac{1}{L}\bra{\Psi_r}\hat \Ep _{as}^\dag(0,t)\hat \Ep _{as}(0,t)\ket{\Psi_r}\approx\nonumber\\
\frac{g^2 L}{c^2}\zeta^2\left(\sum_j e^{-\bar \alpha z_j}e^{-i\Delta^j(t-\tau)}G_j\right)\times\nonumber\\
\left(\sum_\ell e^{-\bar \alpha z_\ell}e^{i\Delta^\ell(t-\tau)}G_\ell^*\right).
\end{align}
Each sum is then replaced by an integral and in the limit where $\theta_0 \ll 1$
\begin{align}
\sum_j e^{-\bar \alpha z_j}e^{-i\Delta^j(t-\tau)}G_j\approx \nonumber\\
\frac{N}{L}\int_0^L dz e^{-\bar \alpha z} \int_{-\infty}^\infty d\Delta \Theta(\Delta) e^{-i\Delta(t-\tau)}.
\end{align}
The spatial integral is carried out directly, while the spectral integral is related to the Fourier transform of the atomic distribution. We obtain
\begin{align}
\frac{1}{L}\bra{\Psi_r}\hat \Ep _{as}^\dag(0,t)&\hat \Ep _{as}(0,t)\ket{\Psi_r}\approx\nonumber\\
\frac{\Gamma}{\sqrt{2\pi} c}&(1-e^{-\bar\alpha L})|\bar\Theta(t-\tau)|^2.
\end{align}
By evaluating the Fourier transform of the chosen atomic distribution (\ref{dist}) for the first revival time, i.e. around $t -\tau= 2\pi/\Delta_0,$ we end up with
\begin{align}
\nonumber
\frac{\sqrt{2\pi}}{\Gamma} \frac{c}{L} \bra{\Psi_r} \hat \Ep _{as}^\dag(0,2\pi/\Delta_0 + \tau)\hat \Ep _{as}(0,2\pi/\Delta_0 + \tau) \ket{\Psi_r} \\
\nonumber
\approx (1-e^{-\bar\alpha L})e^{-\frac{\pi^2}{2 \ln 2\, F^2}}.
\end{align}
The success probability to find an anti-Stokes photon within a temporal mode of duration $\sqrt{2\pi}/\Gamma$ centered at $2\pi/\Delta_0+\tau$ approaches one for large enough optical depth provided that the comb finesse is optimized.\\
Note that for a quantum memory based on atomic frequency comb, the efficiency has a similar expression but the term $(1-e^{-\bar\alpha L})$ is squared because the photon has first to be absorbed before being reemitted. For weak optical depths, a spontaneous Raman scheme is thus more efficient than the corresponding quantum memory. For example, for $\alpha L=0.1$ and optimized finesses, the efficiency of spontaneous Raman scheme reaches 1\% whereas it is limited to 0.1\% for a quantum memory based on atomic frequency comb.  \\
Further note that for an emission in the forward direction, the readout efficiency is given by
\begin{equation}
\frac{(\bar\alpha L)^2 e^{-\bar\alpha L}}{1-e^{-\bar \alpha L}}e^{-\frac{\pi^2}{2 \ln 2\, F^2}}
\end{equation}
and is limited to 65\% by reabsorption instead of 54\% for a quantum memory \cite{Sangouard07}.\\

\paragraph{Noise}
The atoms excited at the write level can produce spontaneous noise in the anti-Stokes mode of interest. To get a lower bound on the signal-to-noise ratio, we assume that all the atoms transfered to the excited state at the write level decays on the e-s transition. The evaluation of the resulting noise is similar to the one associated to the collective emission, excepts that the atomic state is now given by $\varrho_n$ (see eq. (10) in the main text). By applying $\hat \Ep _{as}(0,t)$ on $\rho_n,$ we obtain
\begin{align}
i \frac{gL}{c}\sum_j \theta_0^2 e^{-\frac{3\bar\alpha}{2} z_j} e^{-i\Delta^j(t-(\tau+t_d))}e^{i \omega_{as}(\tau + t_d)}\coh{g_j}{e_j}\times
\nonumber\\
 \bigotimes_{\ell\neq j} ( (1-\theta_0^2 e^{-\bar \alpha z_\ell})\coh{g_\ell}{g_\ell}+ \theta_0^2 e^{-\bar \alpha z_\ell}\coh{e_\ell}{e_\ell} \otimes\coh{0}{0}.
\end{align}
We then apply the adjoint $\hat \Ep_{as}^\dag (0,t)$ before taking the trace. One sees that for the trace to be non-zero, each term $e^{i \omega_{as}(\tau + t_d)}\coh{g_j}{e_j}$ has to be multiplied by the corresponding operator $\sigma_{eg}^j(t+t_d)$ from $\hat \Ep_{as}^\dag (0,t).$ Since the part of the state on the second line has a trace equal to one, we get
\begin{align}
\nonumber \frac{1}{L} \text{tr} \big(\hat \Ep_{as}^{\dagger}(0,2\pi/\Delta_0+\tau) \hat \Ep_{as}(0,  2\pi/\Delta_0+\tau)
\rho_n \big)\\
\nonumber =\frac{g^2L}{c^2}\sum_j \theta_0^2 e^{-2\bar\alpha z_j}
\end{align}
and we conclude that the noise is bounded by $\frac{\theta_0^2}{2}(1-e^{-2 \bar \alpha L})$ by replacing the discrete sum by its integral expression and by taking the duration of the anti-Stokes mode into account. \\
Therefore, the signal-to-noise is bounded by
$$
\text{signal-to-noise} \geq \frac{2(1-e^{-\bar \alpha L})}{\theta_0^2(1-e^{-2\bar\alpha L})}e^{-\frac{\pi^2}{2 \ln 2 \, F^2}}
$$
i.e. for large optical depth and optimized finesse
$$
\text{signal-to-noise} \geq \frac{2}{\theta_0^2}.
$$
Note that in the case of a quantum memory, the photon to be stored is subject to many defective manipulations, e.g. to non-unit coupling efficiency into mono-mode fibers or to imperfect spectral filters. This reduces the signal and hence, strongly limits the signal-to-noise ratio. This is a significant drawback with respect to spontaneous Raman based sources where the readout efficiency is conditioned on the successful detection of a Stokes photon.

\end{document}